\documentclass{article}

\usepackage{arxiv}

\usepackage[utf8]{inputenc} 
\usepackage[T1]{fontenc}    
\usepackage{hyperref}       
\usepackage{url}            
\usepackage{booktabs}       
\usepackage{amsfonts}       
\usepackage{amsmath}
\usepackage{graphicx}       

\title{Thermo-optical dynamics of a nonlinear GaInP photonic crystal nanocavity depend on the optical mode profile}

\author{
    Karindra Perrier, Sebastiaan Greveling, Hilbrand Wouters, Sanli Faez and Allard P. Mosk\\
    Nanophotonics, Debye Institute for Nanomaterials Science \\
    Utrecht University\\ 
    Princetonplein 1, 3584 CC Utrecht, The Netherlands\\
    \texttt{k.l.perrier@uu.nl}\\
    \AND Ga\"{e}lle Lehoucq, Sylvain Combri\'{e} and Alfredo de Rossi\\
    Thales Research and Technology\\
    Route D\'{e}partementale 128, 91767 Palaiseau, France
    \AND Said R. K. Rodriguez\\
    AMOLF Institute, Center for Nanophotonics\\
    Science Park 104, 1098 XG Amsterdam, The Netherlands\\
    }

\begin{document}
\maketitle

%%%%%%%%%%%%%%%%%%% ABSTRACT %%%%%%%%%%%%%%%%

\begin{abstract}
We measure the dynamics of the thermo-optical nonlinearity of both a mode-gap nanocavity and a delocalized mode in a Ga$_{\mathrm{0.51}}$In$_{\mathrm{0.49}}$P photonic crystal membrane. We model these results in terms of heat transport and thermo-optical response in the material.
By step-modulating the optical input power we push the nonlinear resonance to jump between stable branches of its response curve, causing bistable switching. An overshoot of the intensity followed by a relaxation tail is observed upon bistable switching. 
In this way, the thermal relaxation of both the localized resonance and the delocalized resonance is measured. Significant difference in decay time is observed and related to the optical mode profile of the resonance.
We reproduce the observed transient behavior with our thermo-optical model, implementing a non-instantaneous nonlinearity, and taking into account the optical mode profile of the resonance, as experimentally measured. 
\end{abstract}

%%%%%%%%%%%%%%%%%%%%%%%%%%%%%%%%%%%%%%%%%%%%%%%%%%%%%%%%%%%%
%%%%%%%%%%%%%%%%%%%%%%%%%%  body  %%%%%%%%%%%%%%%%%%%%%%%%%%
%%%%%%%%%%%%%%%%%%%%%%%%%%%%%%%%%%%%%%%%%%%%%%%%%%%%%%%%%%%%

%%%%%%%%%%%%%%%%%%%%%%%%%%  1. INTRODUCTION  %%%%%%%%%%%%%%%%%%%%%%%%%%
\section{Introduction}

Resonators on photonic crystal (PhC) membranes are promising candidates for switchable and programmable optical communication devices as well as for fundamental physics exploration, as they combine high field enhancement with strong optical nonlinearity. 
Intentional waveguide defects on these membranes form nanocavities with a high-quality factor (high-Q) and small mode volume~\cite{akahane2003high} that can be connected into coupled-resonator optical waveguides~\cite{Notomi2008,Sokolov:Disorder}. 
The optical nonlinearities in these PhCs are fundamental in the development of all-optical high-bandwidth functionalities like optical switches~\cite{almeida2004all,Tanabe:05,preble2005ultrafast,BruckAndMuskens2015device,Yuce2011ultimate,qiu2017all}, optical memory~\cite{Kuramochi2014}, slow light~\cite{melloni2010tunable,morichetti2012first,Kobus_slowlight,baba2008slow} and quantum networks~\cite{lodahl2017quantum,AtatureAndEnglund2018material}. 
It is known that these nonlinearities can give rise to bistable behavior at powers of only a few $\mu $W~\cite{weidner2007nonlinear,vuckovic2004two}, and the timescale of the dominating nonlinearity will ultimately limit the bandwidth of possible devices. 
The strong optical nonlinearity of PhC cavities can also be used to investigate the physics of dissipative phase transitions~\cite{fitzpatrick2017observation,fink2018signatures,rodriguez2017probing,angerer2017ultralong}, and to probe the interplay of disorder and nonlinearity which is expected to strongly impact transport properties of cavity arrays~\cite{debnath2017nonequilibrium,Hartmann_theoryCROW}. 

Specifically in PhC membranes, the interplay between resonances and their 2D optical mode profiles can produce a strong thermo-optical nonlinearity that enables bistable switching. 
Measurements of the transient behavior brought about by bistable switching is usually controlled with an optical pump, either by pumping a second optical mode~\cite{Notomi:05}, or by using an out-of-plane optical pump pulse~\cite{morsy2019high,brunstein2009thermo, arbabi2012dynamics}. 
However, when there is absorption of the probe the optical mode itself becomes a source of heat, eliminating the need for an optical pump and enabling bistable switching with a single source of drive light.

In this paper, we study the effect of the spatial mode profile on the thermo-optical response time of PhC resonances. 
Thermo-optical bistability induced by resonances have been observed before~\cite{carmon2004dynamical,biswas2015photonic,Gao:17}. Yet, to our best knowledge, an investigation of different resonances involving their optical mode profiles and profile-specific dynamics has not been reported before. 
We measure two modes in a Ga$_{\mathrm{0.51}}$In$_{\mathrm{0.49}}$P (GaInP) PhC slab
by experimentally investigating the temporal behavior in the bistable regime of both a mode-gap cavity resonance and a delocalized waveguide resonance, demonstrating significantly different decay times. 
Additionally, we present a thermo-optical model that takes into account the obtained optical mode profile of the resonance. 
Although many models, formulating the resonator as a single temperature thermal object, have been shown throughout the literature \cite{kim2007all, brunstein2009thermo, carmon2004dynamical, qiu2017all, morsy2019high, ikeda2008thermal}, a model employing the measured mode profile of the resonance has not been demonstrated before to the best of our knowledge.

%%%%%%%%%%%%%%%%  2. PHOTONIC CRYSTAL DESIGN AND EXPERIMENTAL SETUP %%%%%%%%%%%%%%%%%

\section{Characterization of the PhC}\label{sec:characterization}

\begin{figure}[!t]
\centering\includegraphics[width=11.9cm,trim={0 0 3cm 0},clip]{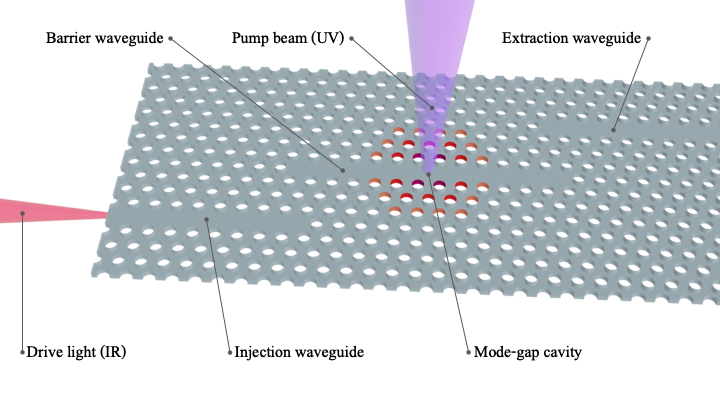}
\caption{Schematic overview of the photonic crystal membrane. 
The mode-gap cavity consists of a broadened waveguide; the air holes lining the waveguide are shifted outwards from their lattice position with $0.0124\sqrt{3} a$, $\frac{2}{3} \times 0.0124\sqrt{3} a$ and $\frac{1}{3} \times 0.0124\sqrt{3} a$ for the first (purple), second (red) and last (orange) row respectively. The cavity is positioned on the barrier waveguide of width $W_0 = 0.98 \sqrt{3} a$. Injection and extraction waveguides, both of width $W_1 = 1.1\sqrt{3} a$, are used to evanescently couple the infrared light in and out.}
\label{fig:sample}
\end{figure}

The air-suspended GaInP membrane, schematically depicted in Fig.~\ref{fig:sample}, is $180$~nm thick with air holes on a hexagonal lattice with a lattice constant of $a = 485$~nm. An injection waveguide is used to evanescently couple infrared light into the barrier waveguide on which the mode-gap cavity is positioned. The mode-gap cavity consists of a local width modulation of the barrier waveguide. In addition to the mode-gap cavity, high-Q delocalized modes spanning the barrier waveguide occur.
The fabrication method is described in detail in Ref.~\cite{Combrie:fabrication}.
The PhC is mounted in a nitrogen filled dust-free box to prevent oxidation. 

The sample is driven with infrared light from a tunable CW laser, injected with a polarization maintaining lensed fiber. 
A 10~GHz bandwidth Mach-Zehnder electro-optic modulator (EOM) controls the input power.
We measure the cavity response using the out-of-plane scattering light collected through a 0.4 NA objective and detect with an InGaAs photodiode (PD) (Hamamatsu G8605-21). An aperture in the conjugate plane removes background scattering. The signal is amplified (Femto dlpca-200) at a bandwidth of 
7~kHz for the mode profiles (Fig.~\ref{fig:modeprofile}) and 
1~kHz, 7~kHz, and 400~kHz for the spectral scans at low input power (Fig.~\ref{fig:spectra}a), high input power (Fig.~\ref{fig:spectra}b), and input power modulation measurements (Fig.~\ref{fig:HystLoop}, Fig.~\ref{fig:overshoot_all} and Fig.~\ref{fig:CompareOvershoot}), respectively.  
An analog-to-digital converter board (NI USB-6356) acquires the spectral scans with a resolution of 1 pm. For the high-bandwidth power modulation measurements we use an oscilloscope at a sample rate of 2~MHz.

In addition to the infrared driving light, a second 405~nm diode laser (OBIS 405LX) provides out-of-plane pump light, focused on the sample by the objective. Due to absorption by the PhC membrane~\cite{sokolov2015local}, the pump applies a local heat source. 
Using a fast scanning mirror we can spatially scan the PhC membrane with the pump and probe the thermo-optical profile of any PhC resonance~\cite{Lian:Profiles}. With this method we identify the modes and measure their profile~\cite{johnson_photonic_2001}.

%%%%%%%%%%%%%%%%%% 2.1 MODE PROFILE %%%%%%%%%%%%%%%%%%%%%%
\subsection{Thermo-optically Measured Mode Profile}
\begin{figure}%[!t]
\centering\includegraphics[width=15.3cm, trim={0.4cm 0 0.4cm 0}, clip]{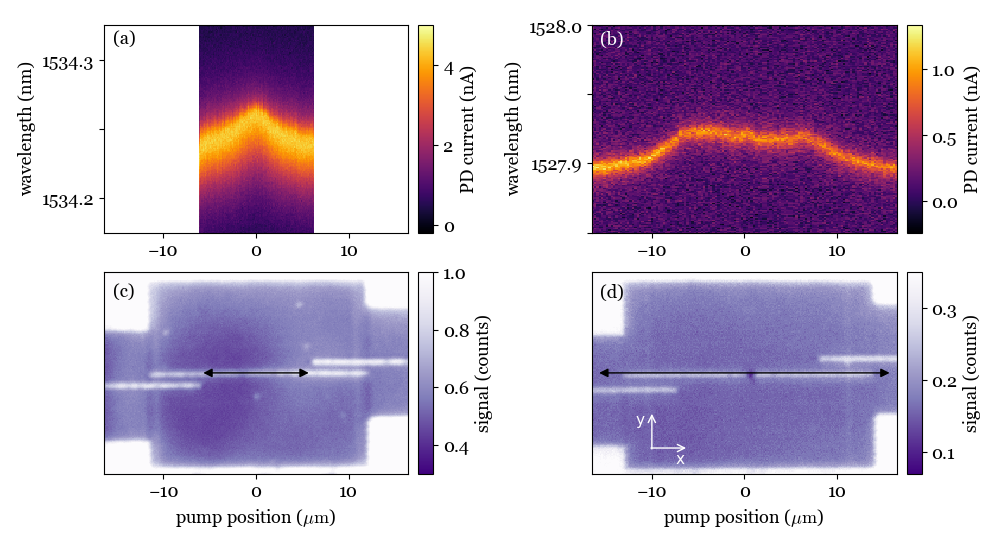}
\caption{Optical mode profile of (a) the localized mode-gap resonance and (b) the delocalized waveguide resonance, obtained with wavelength scans while pumping with a weak pump spot over a range of pump positions on the waveguide. Pump positions are indicated by the black arrows on the image of (c) the PhC membrane holding the mode-gap resonance and (d) the PhC membrane holding the delocalized resonance. The membrane holding the mode-gap resonance has a barrier waveguide length of 24.7~$\mu$m, an overlap between barrier and injection/extraction waveguide of 6.1~$\mu$m in the x-direction and a distance between the waveguides of 1.7~$\mu$m in the y-direction. For the membrane holding the delocalized mode these dimensions are 28.1~$\mu$m, 6.3~$\mu$m and 1.3~$\mu$m respectively.
}
\label{fig:modeprofile}
\end{figure}

In Figs.~\ref{fig:modeprofile}(a-b) we show the thermo-optically measured mode profiles of the two resonances of interest, detected via the out-of-plane scattered light. The corresponding images of the PhC membranes are shown beneath the mode profiles in Figs.~\ref{fig:modeprofile}(c-d). 
The mode profile is measured by scanning the barrier waveguide with the pump, positioning the pump spot in consecutive steps along the \textit{x}-axis of the waveguide. For each pump position a wavelength scan is taken with the drive light. 
Due to the thermo-optical effect of the semiconductor material, local heating induces a redshift of the resonance wavelength when the thermal profile of the pump overlaps with the optical mode. In this way, plotting the wavelength redshift versus pump position reveals the mode profile.

In Fig.~\ref{fig:modeprofile}(a) a range of 12~$\mu$m is scanned in steps of 0.13~$\mu$m with a pump power of 7~$\mu$W. This shows a localized mode positioned at the center of the barrier waveguide with a FWHM of 3~$\mu$m. We thus identify this resonance as a a mode-gap resonance.
Fig.~\ref{fig:modeprofile}(b) is measured over a wider pump range of 33~$\mu$m, taken with pump position steps of 0.26~$\mu$m and a pump power of 6~$\mu$W. This shows a profile that spans the barrier waveguide between the ends of the injection and extraction waveguide, with a maximum redshift over a 14-$\mu$m range. We therefore identify this resonance as a delocalized mode.

%%%%%%%%%%%%%%%%%% 2.2 SPECTRAL SCAN %%%%%%%%%%%%%%%%%%%%%%
\subsection{Spectral Scan} \label{sec:Spectral Response}

\begin{figure}[!t]
\centering\includegraphics[width=13.0cm, trim={-0.4cm 0 0.0cm 0}, clip]{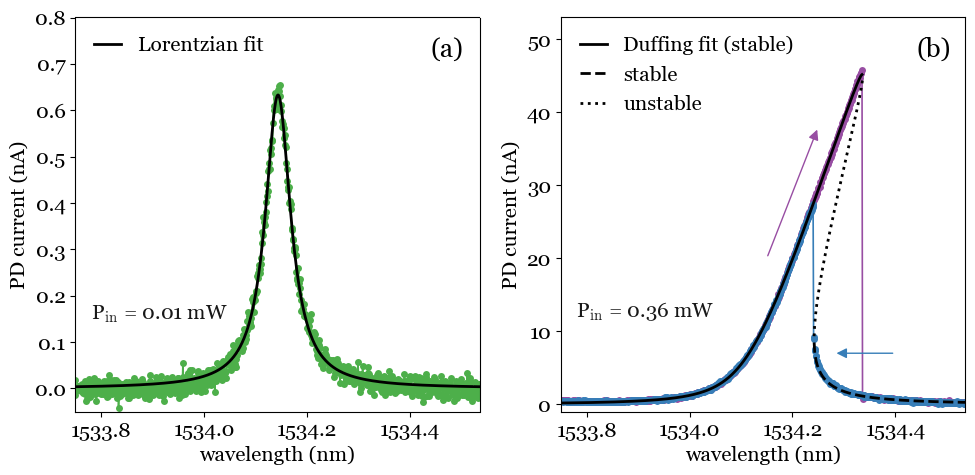}
\caption{Spectral response of the mode-gap resonance in the (a) linear regime and (b) nonlinear regime as measured via the out-of-plane scattered light. In the nonlinear regime, the response depends on the wavelength scan direction, as indicated by the arrows for the forward (purple) and backward (blue) scans.
For the nonlinear response we fit the Duffing equation to the upper steady-state branch (solid black curve) to find nonlinear parameter $\beta$ and then plot the corresponding solutions for the lower steady-state branch (dashed) and the unstable branch (dotted).
Scan rates were selected to obtain a good signal to noise ratio, with 0.5 nm/s for the low intensity linear regime and 1 nm/s for the high intensity nonlinear regime.
}
\label{fig:spectra}
\end{figure}

In Fig.~\ref{fig:spectra} we show the spectral response of the mode-gap nanocavity measured via the out-of-plane scattered intensity. Spectral scans of the delocalized mode show a similar response and are therefore not depicted.
Fig.~\ref{fig:spectra}(a) shows the spectral response at low input power demonstrating the lineshape in the linear response regime. A Lorentzian lineshape fit yields the bare resonance $\lambda_0 = 1534.144$~nm and cold cavity linewidth $\Gamma = 30$~pm. For the delocalized mode we find a bare resonance of $\lambda_0 = 1527.736$~nm and linewidth of $\Gamma = 11$~pm in the same way. 

In Fig.~\ref{fig:spectra}(b) we show the spectrum in the nonlinear regime with a bidirectional scan at high input power. The nonlinear behavior is characterized by a resonance peak that is skewed towards longer wavelengths, exhibiting a bistable response that depends on the direction of the scan. The forward scan follows the upper steady-state branch until, at large detuning, the system reaches the end of the branch and jumps down discontinuously, demarcating the edge of the bistable regime. For the backward scan the system follows the lower steady-state branch, undercutting the upper branch, until it reaches the other edge of bistability and jumps back up. In this way a hysteresis loop that easily spans $100$~pm is formed, demonstrating strong nonlinearity in the cavity. 

To fit the spectral response in the nonlinear regime we solve the Duffing equation~\cite{Duffing-Ueda}, describing a resonator with an instantaneous third-order nonlinearity, using the $\lambda_0$ found in Fig.~\ref{fig:spectra}(a). 
The equation will return both stable and unstable solutions that form a tilted peak with an upper and lower branch that are connected by the unstable branch. We fit the upper steady-state branch and use the result to plot the solutions for both the lower branch and the unstable branch.
The fit yields the linewidth of the hot cavity, heated by absorption near resonance, $\Gamma_{\mathrm{hot}} = 24$~pm, and a negative nonlinear parameter, which is in accordance with a redshift of the resonance in the bistable regime. We keep the hot cavity linewidth a free fit parameter because the temperature affects the resonance frequency, and indirectly through that, the mode profile and the losses to the injection and extraction waveguides~\cite{LianDispersion, faggiani2016lower}. For the delocalized mode we find qualitatively the same nonlinear behavior with a negative nonlinear parameter, except here we have a slightly broader heated linewidth of $\Gamma_{\mathrm{hot}} = 14$~pm.

%%%%%%%%%%%%%%%%%%%%%%%% 2.2. INPUT POWER SCAN %%%%%%%%%%%%%%%%%%%%%%%%%
\subsection{Input Power Scan}

\begin{figure}[!t]
\centering\includegraphics[width=12.6cm,trim={1.5cm 0 2cm 0}, clip]{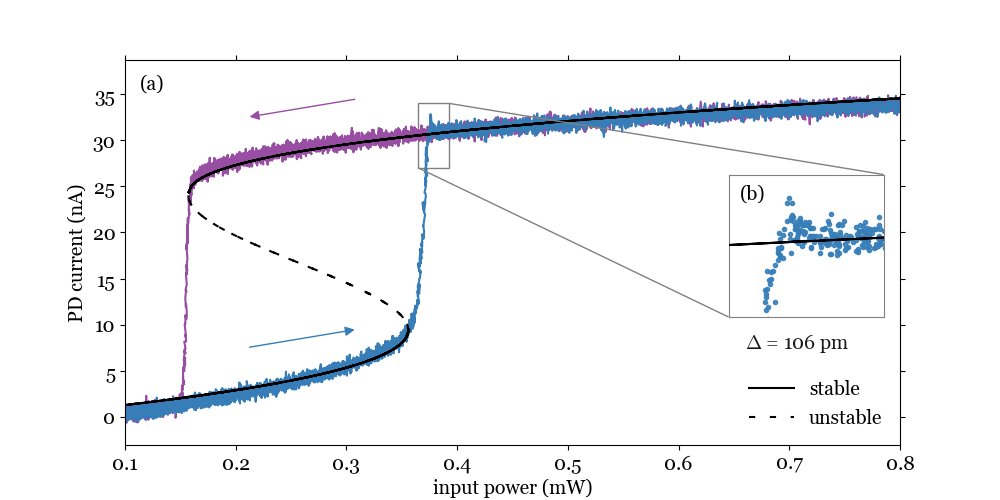}
\caption{(a) Hysteresis loop in the nonlinear regime of the mode-gap resonance for fixed detuning $\Delta = 106$~pm and a 100~Hz triangle wave input power modulation. Arrows indicate the power scan direction. 
The out-of-plane scattered light is averaged over 100 modulation periods.
Solutions for the Duffing equation are plotted for the both stable and unstable solutions as a guide to the eye. The inset (b) shows the details of an overshoot of the upper steady-state branch.}
\label{fig:HystLoop}
\end{figure}

To characterize the hysteresis behavior we measure out-of-plane scattering light versus input power. In Fig.~\ref{fig:HystLoop}(a) we scan the input power for the mode-gap resonance with a 100~Hz triangular wave while keeping the wavelength fixed at a positive detuning of $\Delta \equiv \lambda - \lambda_0 = $ 106~pm. The measured out-of-plane scattering light is averaged over 100 modulation periods.
The resulting cavity response shows the hysteresis loop: a region of bistability with discontinuous jumps at the ends of the upper and lower steady-state branch. 

As a guide to the eye we plot the steady-state solutions of the Duffing equation versus input power, using the parameters found by the spectral fits of Section \ref{sec:Spectral Response}, using $\Gamma$ for both the lower branch and unstable branch and $\Gamma_{\mathrm{hot}}$ for the upper branch. As the bare cavity resonance slightly drifts, we adjust $\lambda_0$ found in Section~\ref{sec:Spectral Response} to correct for drift between and during scans, with a maximum correction of 42~pm. We attribute this slow shift to a thin water layer on the membrane, as explained elsewhere~\cite{H2Oasay2005,chen2011selective,sokolovThesis,zhang2014dissociative}, and corresponds to the evaporation of a <1~nm thick water layer. The S-shaped steady-state solutions of the Duffing equation are qualitatively consistent with the data -- predicting both stable steady-state branches and the loop edges at the unstable turning points -- except for the observation of an intensity overshoot. 

Fig.~\ref{fig:HystLoop}(b) shows this intensity overshoot in detail. It is induced at the moment the system reaches the end of the lower steady-state branch and is forced into the upper steady-state branch. 
We perform a control measurement to rule out the overshoot being an artifact of the apparatus. With a step response measurement of the EOM-amplifier-oscilloscope chain a clean $10\%-90\%$ rise time of $1.5\,\mu$s is observed, showing no overshoot. 
Since the Duffing equation describes an oscillator with instantaneous nonlinearity, it cannot describe the dynamics of the thermo-optical nonlinearity causing the overshoot. An adequate model is presented in Section~\ref{sec:ThermalModel}.

We note that the parameters we chose for our experiments are such that our hysteresis measurements can be compared with a stationary solution~\cite{rodriguez2017probing}. Two considerations are made, the relevant timescale of the spectral and power scans are longer than the thermo-optical relaxation time. Furthermore, the scan timescale is short compared to the characteristic time at which fluctuations induce stochastic switching between bistable states.
The input power scans of the delocalized mode result in similar hysteresis loops, with an even more pronounced overshoot of the upper steady-state branch.

\subsection{Time-resolved Measurement of Thermo-optical Dynamics}\label{sec:OvershootDetunings}

\begin{figure}[!t]
\centering\includegraphics[width=12.3cm]{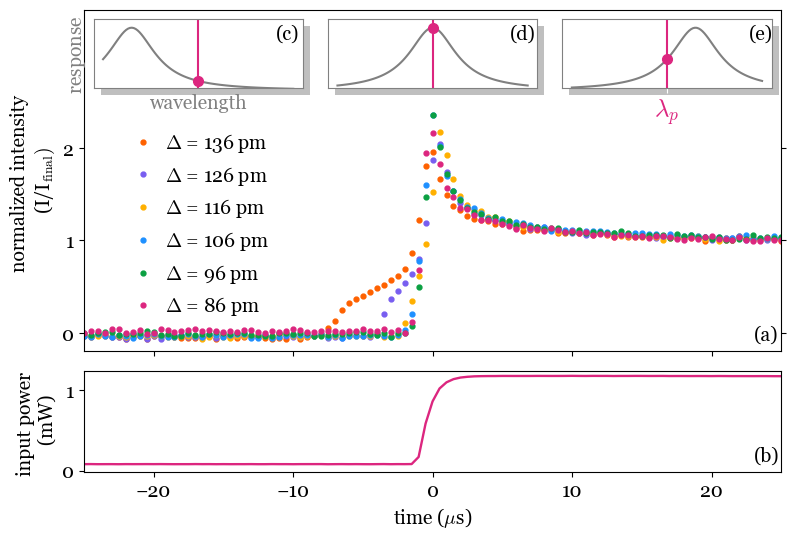}
\caption{(a) Relaxation of the thermal overshoot at six different detunings measured in the out-of-plane scattered intensity. (b) Input power used to obtain the overshoot at $\Delta = 86$~pm. All other relaxation curves have time offsets between 0 and -8~$\mu$s such that the peak intensities are aligned. (c-e) Illustration of a typical instantaneous cavity response at the three different stages of the power step: (c) at the initial steady-state before the overshoot when we have low input power, (d) at the moment of jumping states when the input power steps up, (e) at the final steady-state when we have high input power.  
}
\label{fig:overshoot_all}
\end{figure}

In Fig.~\ref{fig:overshoot_all}(a) we show the time-resolved relaxation of the intensity overshoot of the mode-gap resonance measured via the out-of-plane scattered power, averaged over 100 cycles. 
The overshoot is obtained by step-modulating the input power with a lead edge of 1.5~$\mu$s, as depicted in Fig.~\ref{fig:overshoot_all}(b). 
In this way we drive the system across the hysteresis loop beyond the end of the lower steady-state branch and thus force a jump to the upper steady-state branch. 
We do this for six different detunings where each measurement is aligned at the peak intensity, showing the relaxation behavior right after the peak of the overshoot is independent of detuning. 
The emerging plateau preceding the overshoot for large detunings is the only part of the behavior that is affected by varying the detuning. The increasing duration of the plateau for larger detunings is behavior we have seen in computations with our model described in Section~\ref{sec:ThermalModel}. However, a detailed discussion of the physics preceding the overshoot lies beyond the scope of this paper. 
Note that the timescale of the overshoot is too fast to be detected in the spectral scans, and therefore could only be observed in the high-bandwidth power sweep measurements. 

In Figs.~\ref{fig:overshoot_all}(c-e) we illustrate the instantaneous cavity resonance corresponding to different moments in the power step to explain the physical mechanism of the intensity overshoot, as it helps to separate the resonant behavior of the cavity from its thermo-optical shift. 
At the start of the scan the cold cavity resonance lies on the blue side of the driving wavelength (Fig.~\ref{fig:overshoot_all}(c)). When the driving power is increased, the resonance wavelength will shift towards the red due to absorption of the drive light. This \emph{pulls} the system through resonance (Fig.~\ref{fig:overshoot_all}(d)), momentarily maximizing the intensity inside the cavity. This is manifested in the sharp intensity peak. The resonance shifts further to the red due to more absorption until the cavity reaches its steady state temperature. The hot cavity resonance is now on the red side of the driving wavelength, stabilized in the upper steady-state branch (Fig.~\ref{fig:overshoot_all}(e)). 
A similar non-instantaneous response is expected when the system falls from the upper steady-state branch onto the lower one while sweeping in the opposite direction.

%%%%%%%%%%%%%%%%%%%%%% 3. THERMO-OPTICAL MODEL %%%%%%%%%%%%%%%%%%%%%%%
\section{Thermo-optical Model}\label{sec:ThermalModel}

In this section we will focus on the dynamics of the transient. We consider the relaxation of the overshoot including the peak, meaning the stages of the power step depicted in Figs.~\ref{fig:overshoot_all}(d,e). 
To describe the resonance shifts and the thermal diffusion in the PhC membrane, we solve a set of coupled equations relating the optical resonance to a modified heat equation.
The electromagnetic response of the cavity is assumed to be on the timescale of the cavity photon lifetime of order 100~ps. Therefore, the electromagnetic response is effectively decoupled from the thermal diffusion in the PhC, which happens on a timescale of order 100~ns.
Neglecting the \textit{z}-dependence in the thin semiconductor slab, the optical energy density inside the cavity is
\begin{equation}
    U_{\mathrm{opt}}(x,y,t)= \tau_{\mathrm{in}} P_{\mathrm{in}}(t)  U_{\mathrm{mode}}(x,y) 
    \frac{\Gamma^2}{\Gamma^2+(\Delta-\delta_{\mathrm{th}}(t) )^2} ,
    \label{eq:resonancecondition}
\end{equation}

\noindent
where $\tau_{\mathrm{in}}$ is the input coupling efficiency, $P_{\mathrm{in}}$ is the input power of the drive light, 
$U_{\mathrm{mode}}$ is the energy density of the optical mode profile, 
and $\delta_{\mathrm{th}}(t)$ is the thermal resonance shift that causes a dynamic detuning.

The energy density of the optical mode is normalized over the \textit{xy}-plane of the membrane
\begin{equation}
    \int U_{\mathrm{mode}}(x,y) dx dy = 1.
\end{equation}

\noindent
For our experimental conditions the resonance frequency shifts linearly with temperature~\cite{Sokolov:Thermo-opticalCoeff}. The thermal resonance shift is then given by
\begin{equation}
    \delta_{\mathrm{th}}(t)= \int \eta U_{\mathrm{mode}}(x,y)  T(x,y,t) dx dy,
\end{equation}
where $\eta$ is the thermo-optical coefficient of the semiconductor material and $T$ the temperature profile.
We consider the two cooling processes that dissipate heat into the substrate: in-plane heat dissipation that eventually reaches the bridge connecting the PhC with the substrate, and dissipation via the thin gas layer separating the membrane and the substrate beneath the PhC. Conductance through the gas on the top side of the membrane is neglected as it is likely to represent only a small fraction of the dissipation.
Then, the dynamics of the temperature profile are described by the heat equation,

\begin{equation}
   C_{\mathrm{2D}} \frac{\partial T(x,y,t)}{\partial t} = - K_{\mathrm{2D}} \nabla^2 T(x,y,t) - K_{\mathrm{gas}} (T(x,y,t)-T_0) + \alpha\Gamma U_{\mathrm{opt}}(x,y,t),
   \label{eq:heatequation}
\end{equation}
with the boundary condition that $T=T_0$ at the edges of the PhC membrane where there is a high thermal conductivity link with the substrate. 
Here, $C_{\mathrm{2D}}$ is the 2D specific heat of the PhC membrane, $K_{\mathrm{2D}}$ is the 2D thermal conductivity of the PhC membrane, $K_{\mathrm{gas}}$ is the thermal conductance of the gas layer between the PhC membrane and the substrate, and $\alpha$ the absorption fraction, i.e., the probability a cavity photon is absorbed rather than scattered or leaked out of the cavity.

The 2D specific heat is given by 
\begin{equation}
    C_{\mathrm{2D}}= h \phi \rho C_{\mathrm{sp}},
\end{equation}
with $h$ the slab thickness of the PhC membrane, $\phi$ the filling fraction of the PhC, $\rho$ the density and $C_{\mathrm{sp}}$ the specific heat of the semiconductor material.
The thermal conductivity $K_{\mathrm{2D}}$ is approximated as $h \phi \kappa$ where $\kappa$ is the thermal conductivity of the semiconductor. Out-of-plane conductance through the gas layer is modeled by $K_{\mathrm{gas}}=\kappa_{\mathrm{gas}}/d_{\mathrm{gas}}$, where $d_{\mathrm{gas}}$ is the distance between the membrane and the substrate. The contribution of the gas to the in-plane thermal conductance is neglected, as it is much smaller than the membrane conductance.

The system of equations, composed of Eq.~\eqref{eq:heatequation}, the heat equation which is a partial differential equation, and Eq.~\eqref{eq:resonancecondition}, the resonance condition, is solved using Euler's method~\cite{schwarz1989numerical} on an $(x,y,t)$ grid. While not efficient, Euler integration is extremely robust as long as the step sizes fulfill the appropriate Courant-Friedrichs-Lewy condition~\cite{courant1928partiellen}.

%%%%%%%%%%%%%%%%%% Dynamics of Different Mode Profiles %%%%%%%%%%%%%%%%%%%
\section{Dynamics of Different Mode Profiles}\label{sec:ComparingOvershoot}

\begin{figure}[t!]
\centering\includegraphics[width = 13.0cm]{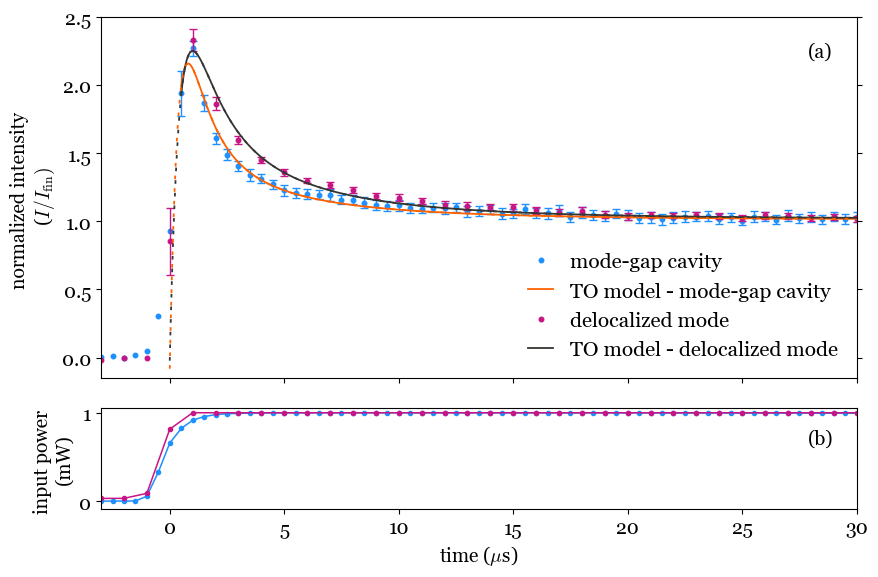}
\caption{(a) The transient behavior upon bistable switching showing thermal relaxation of both the mode-gap resonance and the delocalized resonance, at small but positive detunings of $3.5\Gamma$ and $1\Gamma$ respectively. The response was measured via the out-of-plane scattered intensity, obtained by step-modulating the input power as shown in (b).
Theoretical fits are plotted in solid lines with dashed curves in the extrapolation regions. Error bars are indicated for both data sets.
}
\label{fig:CompareOvershoot}
\end{figure}

In Fig.~\ref{fig:CompareOvershoot} we show the response of the two different modes shown in Fig.~\ref{fig:modeprofile}. Fig.~\ref{fig:CompareOvershoot}(a) shows the measured relaxation behavior of the intensity overshoot of both modes in response to the step-modulated input power shown in Fig.~\ref{fig:CompareOvershoot}(b). 
The delocalized mode has a longer 90\%$-$10\% decay time of $\tau_{90-10}=$ 10.1$\pm$1.8$\,\mu$s compared to the mode-gap cavity with $\tau_{90-10}=$ 7.7$\pm$2.2$\,\mu$s, where the 3$\sigma$ error limits have been estimated by considering the spread of the decay times observed in different subsets of our data. In high-thermal conductivity platforms, like Si~\cite{Notomi:05,Yu:20} or GaAs~\cite{de2009interplay}, the thermo-optics are typically much faster.

We compare the experimental data to our thermo-optical model and do a fit on both modes, using the appropriate measured mode profile $U_{\mathrm{mode}}$ in each case. The fitted model is shown in Fig.~\ref{fig:CompareOvershoot}(a), using the specific heat of GaInP $C_{\mathrm{sp}} = 310$~J/kgK~\cite{Piesbergen:SpecificHeatInP}, GaInP density $\rho = 4.81\times 10^3$~$\mathrm{kg/m^3}$~\cite{density}, a fill fraction of 0.714 for the membrane to account for the air holes, and an estimated value of $\alpha = 0.1$ for the absorption fraction. This absorption fraction is the highest value that is compatible with the sharpest observed resonances in these samples. We stress that the model is only sensitive to the product of the three parameters $\alpha\tau_{\mathrm{in}} P_{\mathrm{in}}$. Hence the fit does not depend on our estimate for $\alpha$. 

Since we focus on the dynamics starting around the peak of the overshoot, we start the fit one data point before the peak intensity. The detuning we fix at $\Delta = 0$ to assure we start modeling from the moment the drive light is at resonance with the mode (see Fig.~\ref{fig:overshoot_all}(d)). 
We fit three parameters: the input energy $\tau_{\mathrm{in}}P_{\mathrm{in}}$ which is the electromagnetic energy inside the cavity on resonance, a scaling factor relating the intracavity power to the normalized detected out-of-plane scattering light, and an offset parameter accounting for the bias current of the photodiode. For the input energy we find 0.5~fJ for the mode-gap resonance and 1.5~fJ for the delocalized mode. 
These values lie within reasonably expected input powers and incoupling rates. 
We conclude that the delocalized mode, existing in a thermally identical environment as the localized mode, shows significantly slower dynamics. Our model, using the measured mode profiles as input, reproduce the behavior accurately.

%%%%%%%%%%%%%%%%%%%%%%% Discussion %%%%%%%%%%%%%%%%%%%%%%%%%
\section{Discussion}
To understand the difference in decay time we look at which cooling processes dominate.
Both mode profiles have negligible overlap with the edges of the PhC, 
therefore we assume we are in the regime where virtually all of the cooling happens via the gas.
This is confirmed by our model, as for both fitted curves more than 99.9\% of the heat is dissipated via the gas layer.
Hence the dimensions of the membrane and the distances to heat sinks such as the bridge to the substrate play no role here, unlike in the case of more highly thermally conductive materials like Si and GaAs crystals~\cite{weidner2007nonlinear,morsy2019high,brunstein2009thermo,arbabi2012dynamics}.
We note that the larger delocalized resonance has approximately two times the effective volume to surface ratio, and therefore a larger ratio of heat capacity to heat loss rate. 
This explains its longer relaxation time and slower dynamics.

%%%%%%%%%%%%%%%%%%%%%%% Conclusion %%%%%%%%%%%%%%%%%%%%%%%%%
\section{Conclusion}
In conclusion, we measured the response of two different modes in a PhC membrane by measuring the transient behavior upon bistable switching. 
We found that for the localized mode the thermal relaxation is 1.3 times faster than for a delocalized, elongated mode. The experimentally obtained thermo-optical dynamics are modeled using a heat equation that takes into account the optical mode profile of the resonance. 
This showed that, in the regime where heat dissipation predominantly happens through gas cooling via the PhC surface, the aspect ratio of the mode profile determines the resonance's thermo-optical relaxation time.

The influence of the mode profile on the temporal behavior is relevant also in other semiconductor materials, as the mode profile can be used to optimize optical switches for speed or stability.
Targeted heat sinks customized to serve specific modes, designing resonances with different decay times, can serve a number of applications. Different time scales available in one PhC combined with nonlinearities of other origins, like charge-carrier density or the Kerr effect, open the doors to excitability \cite{yacomotti2006fast} and constitutes to tunable all-optical delay lines.

\section*{Acknowledgments}
The authors would like to thank Dante Killian, Cees de Kok, Paul Jurrius and Aquiles Carattino for their technical support. We would also like to thank Dries van Oosten for the helpful discussions. 

\section*{Funding}
This research was funded by Nederlandse Organisatie voor Wetenschappelijk Onderzoek (VENI 189039, VICI 68047618).

\section*{Disclosure}
The authors declare no conflicts of interest.

%%%%%%%%%%%%%%%%%%%%%%% References %%%%%%%%%%%%%%%%%%%%%%%%%
\bibliographystyle{unsrt} 
\bibliography{References}

\end{document}